\documentstyle[11pt,paspconf]{article}
 
\def\PsfigVersion{1.10}
\def\setDriver{\DvipsDriver} 
\ifx\undefined\psfig\else\endinput\fi
\let\LaTeXAtSign=\@
\let\@=\relax
\edef\psfigRestoreAt{\catcode`\@=\number\catcode`@\relax}
\catcode`\@=11\relax
\newwrite\@unused
\def\ps@typeout#1{{\let\protect\string\immediate\write\@unused{#1}}}

\def\DvipsDriver{
	\ps@typeout{psfig/tex \PsfigVersion -dvips}
\def\PsfigSpecials{\DvipsSpecials} 	\def\ps@dir{/}
\def\ps@predir{} }
\def\OzTeXDriver{
	\ps@typeout{psfig/tex \PsfigVersion -oztex}
	\def\PsfigSpecials{\OzTeXSpecials}
	\def\ps@dir{:}
	\def\ps@predir{:}
	\catcode`\^^J=5
}

\def\figurepath{./:}

\def\DoPaths#1{\expandafter\EachPath#1\stoplist}
\def\leer{}
\def\EachPath#1:#2\stoplist{
  \ExistsFile{#1}{\SearchedFile}
  \ifx#2\leer
  \else
    \expandafter\EachPath#2\stoplist
  \fi}
\def\ps@dir{/}
\def\ExistsFile#1#2{%
   \openin1=\ps@predir#1\ps@dir#2
   \ifeof1
       \closein1
   \else
       \closein1
        \ifx\ps@founddir\leer
           \edef\ps@founddir{#1}
        \fi
   \fi}
\def\get@dir#1{%
  \def\ps@founddir{}
  \def\SearchedFile{#1}
  \DoPaths\figurepath
}

\def\@nnil{\@nil}
\def\@empty{}
\def\@psdonoop#1\@@#2#3{}
\def\@psdo#1:=#2\do#3{\edef\@psdotmp{#2}\ifx\@psdotmp\@empty \else
    \expandafter\@psdoloop#2,\@nil,\@nil\@@#1{#3}\fi}
\def\@psdoloop#1,#2,#3\@@#4#5{\def#4{#1}\ifx #4\@nnil \else
       #5\def#4{#2}\ifx #4\@nnil \else#5\@ipsdoloop #3\@@#4{#5}\fi\fi}
\def\@ipsdoloop#1,#2\@@#3#4{\def#3{#1}\ifx #3\@nnil 
       \let\@nextwhile=\@psdonoop \else
      #4\relax\let\@nextwhile=\@ipsdoloop\fi\@nextwhile#2\@@#3{#4}}
\def\@tpsdo#1:=#2\do#3{\xdef\@psdotmp{#2}\ifx\@psdotmp\@empty \else
    \@tpsdoloop#2\@nil\@nil\@@#1{#3}\fi}
\def\@tpsdoloop#1#2\@@#3#4{\def#3{#1}\ifx #3\@nnil 
       \let\@nextwhile=\@psdonoop \else
      #4\relax\let\@nextwhile=\@tpsdoloop\fi\@nextwhile#2\@@#3{#4}}
\ifx\undefined\fbox
\newdimen\fboxrule
\newdimen\fboxsep
\newdimen\ps@tempdima
\newbox\ps@tempboxa
\long\def\fbox#1{\leavevmode\setbox\ps@tempboxa\hbox{#1}\ps@tempdima\fboxrule
    \advance\ps@tempdima \fboxsep \advance\ps@tempdima \dp\ps@tempboxa
   \hbox{\lower \ps@tempdima\hbox
  {\vbox{\hrule height \fboxrule
          \hbox{\vrule width \fboxrule \hskip\fboxsep
          \vbox{\vskip\fboxsep \box\ps@tempboxa\vskip\fboxsep}\hskip 
                 \fboxsep\vrule width \fboxrule}
                 \hrule height \fboxrule}}}}
\fi
\newread\ps@stream
\newif\ifnot@eof       
\newif\if@noisy        
\newif\if@atend        
\newif\if@psfile       
{\catcode`\%=12\global\gdef\epsf@start{
\def\epsf@PS{PS}
\def\epsf@getbb#1{%
\openin\ps@stream=\ps@predir#1
\ifeof\ps@stream\ps@typeout{Error, File #1 not found}\else
   {\not@eoftrue \chardef\other=12
    \def\do##1{\catcode`##1=\other}\dospecials \catcode`\ =10
    \loop
       \if@psfile
	  \read\ps@stream to \epsf@fileline
       \else{
	  \obeyspaces
          \read\ps@stream to \epsf@tmp\global\let\epsf@fileline\epsf@tmp}
       \fi
       \ifeof\ps@stream\not@eoffalse\else
       \if@psfile\else
       \expandafter\epsf@test\epsf@fileline:. \\%
       \fi
          \expandafter\epsf@aux\epsf@fileline:. \\%
       \fi
   \ifnot@eof\repeat
   }\closein\ps@stream\fi}%
\long\def\epsf@test#1#2#3:#4\\{\def\epsf@testit{#1#2}
			\ifx\epsf@testit\epsf@start\else
\ps@typeout{Warning! File does not start with `\epsf@start'.  It may not be a PostScript file.}
			\fi
			\@psfiletrue} 
{\catcode`\%=12\global\let\epsf@percent=
\long\def\epsf@aux#1#2:#3\\{\ifx#1\epsf@percent
   \def\epsf@testit{#2}\ifx\epsf@testit\epsf@bblit
	\@atendfalse
        \epsf@atend #3 . \\%
	\if@atend	
	   \if@verbose{
		\ps@typeout{psfig: found `(atend)'; continuing search}
	   }\fi
        \else
        \epsf@grab #3 . . . \\%
        \not@eoffalse
        \global\no@bbfalse
        \fi
   \fi\fi}%
\def\epsf@grab #1 #2 #3 #4 #5\\{%
   \global\def\epsf@llx{#1}\ifx\epsf@llx\empty
      \epsf@grab #2 #3 #4 #5 .\\\else
   \global\def\epsf@lly{#2}%
   \global\def\epsf@urx{#3}\global\def\epsf@ury{#4}\fi}%
\def\epsf@atendlit{(atend)} 
\def\epsf@atend #1 #2 #3\\{%
   \def\epsf@tmp{#1}\ifx\epsf@tmp\empty
      \epsf@atend #2 #3 .\\\else
   \ifx\epsf@tmp\epsf@atendlit\@atendtrue\fi\fi}

\chardef\psletter = 11 
\chardef\other = 12

\newif \ifdebug 
\newif\ifc@mpute 
\c@mputetrue 

\let\then = \relax
\def\r@dian{pt }
\let\r@dians = \r@dian
\let\dimensionless@nit = \r@dian
\let\dimensionless@nits = \dimensionless@nit
\def\internal@nit{sp }
\let\internal@nits = \internal@nit
\newif\ifstillc@nverging
\def \Mess@ge #1{\ifdebug \then \message {#1} \fi}

{ 
	\catcode `\@ = \psletter
	\gdef \nodimen {\expandafter \n@dimen \the \dimen}
	\gdef \term #1 #2 #3%
	       {\edef \t@ {\the #1}
		\edef \t@@ {\expandafter \n@dimen \the #2\r@dian}%
		\t@rm {\t@} {\t@@} {#3}%
	       }
	\gdef \t@rm #1 #2 #3%
	       {{%
		\count 0 = 0
		\dimen 0 = 1 \dimensionless@nit
		\dimen 2 = #2\relax
		\Mess@ge {Calculating term #1 of \nodimen 2}%
		\loop
		\ifnum	\count 0 < #1
		\then	\advance \count 0 by 1
			\Mess@ge {Iteration \the \count 0 \space}%
			\Multiply \dimen 0 by {\dimen 2}%
			\Mess@ge {After multiplication, term = \nodimen 0}%
			\Divide \dimen 0 by {\count 0}%
			\Mess@ge {After division, term = \nodimen 0}%
		\repeat
		\Mess@ge {Final value for term #1 of 
				\nodimen 2 \space is \nodimen 0}%
		\xdef \Term {#3 = \nodimen 0 \r@dians}%
		\aftergroup \Term
	       }}
	\catcode `\p = \other
	\catcode `\t = \other
	\gdef \n@dimen #1pt{#1} 
}

\def \Divide #1by #2{\divide #1 by #2} 

\def \Multiply #1by #2
       {{
	\count 0 = #1\relax
	\count 2 = #2\relax
	\count 4 = 65536
	\Mess@ge {Before scaling, count 0 = \the \count 0 \space and
			count 2 = \the \count 2}%
	\ifnum	\count 0 > 32767 
	\then	\divide \count 0 by 4
		\divide \count 4 by 4
	\else	\ifnum	\count 0 < -32767
		\then	\divide \count 0 by 4
			\divide \count 4 by 4
		\else
		\fi
	\fi
	\ifnum	\count 2 > 32767 
	\then	\divide \count 2 by 4
		\divide \count 4 by 4
	\else	\ifnum	\count 2 < -32767
		\then	\divide \count 2 by 4
			\divide \count 4 by 4
		\else
		\fi
	\fi
	\multiply \count 0 by \count 2
	\divide \count 0 by \count 4
	\xdef \product {#1 = \the \count 0 \internal@nits}%
	\aftergroup \product
       }}

\def\r@duce{\ifdim\dimen0 > 90\r@dian \then   
		\multiply\dimen0 by -1
		\advance\dimen0 by 180\r@dian
		\r@duce
	    \else \ifdim\dimen0 < -90\r@dian \then  
		\advance\dimen0 by 360\r@dian
		\r@duce
		\fi
	    \fi}

\def\Sine#1%
       {{%
	\dimen 0 = #1 \r@dian
	\r@duce
	\ifdim\dimen0 = -90\r@dian \then
	   \dimen4 = -1\r@dian
	   \c@mputefalse
	\fi
	\ifdim\dimen0 = 90\r@dian \then
	   \dimen4 = 1\r@dian
	   \c@mputefalse
	\fi
	\ifdim\dimen0 = 0\r@dian \then
	   \dimen4 = 0\r@dian
	   \c@mputefalse
	\fi
	\ifc@mpute \then
		\divide\dimen0 by 180
		\dimen0=3.141592654\dimen0
		\dimen 2 = 3.1415926535897963\r@dian 
		\divide\dimen 2 by 2 
		\Mess@ge {Sin: calculating Sin of \nodimen 0}%
		\count 0 = 1 
		\dimen 2 = 1 \r@dian 
		\dimen 4 = 0 \r@dian 
		\loop
			\ifnum	\dimen 2 = 0 
			\then	\stillc@nvergingfalse 
			\else	\stillc@nvergingtrue
			\fi
			\ifstillc@nverging 
			\then	\term {\count 0} {\dimen 0} {\dimen 2}%
				\advance \count 0 by 2
				\count 2 = \count 0
				\divide \count 2 by 2
				\ifodd	\count 2 
				\then	\advance \dimen 4 by \dimen 2
				\else	\advance \dimen 4 by -\dimen 2
				\fi
		\repeat
	\fi		
			\xdef \sine {\nodimen 4}%
       }}

\def\Cosine#1{\ifx\sine\UnDefined\edef\Savesine{\relax}\else
		             \edef\Savesine{\sine}\fi
	{\dimen0=#1\r@dian\advance\dimen0 by 90\r@dian
	 \Sine{\nodimen 0}
	 \xdef\cosine{\sine}
	 \xdef\sine{\Savesine}}}	      

\def\psdraft{
	\def\@psdraft{0}
}
\def\psfull{
	\def\@psdraft{100}
}

\psfull 

\newif\if@scalefirst
\def\psscalefirst{\@scalefirsttrue}
\def\psrotatefirst{\@scalefirstfalse}
\psrotatefirst

\newif\if@draftbox
\def\psnodraftbox{
	\@draftboxfalse
}
\def\psdraftbox{
	\@draftboxtrue
}
\@draftboxtrue

\newif\if@prologfile
\newif\if@postlogfile
\def\pssilent{
	\@noisyfalse
}
\def\psnoisy{
	\@noisytrue
}
\psnoisy
\newif\if@bbllx
\newif\if@bblly
\newif\if@bburx
\newif\if@bbury
\newif\if@height
\newif\if@width
\newif\if@rheight
\newif\if@rwidth
\newif\if@angle
\newif\if@clip
\newif\if@verbose
\def\@p@@sclip#1{\@cliptrue}
\newif\if@decmpr
\def\@p@@sfigure#1{\def\@p@sfile{null}\def\@p@sbbfile{null}\@decmprfalse
   \openin1=\ps@predir#1
   \ifeof1
	\closein1
	\get@dir{#1}
	\ifx\ps@founddir\leer
		\openin1=\ps@predir#1.bb
		\ifeof1
			\closein1
			\get@dir{#1.bb}
			\ifx\ps@founddir\leer
				\ps@typeout{Can't find #1 in \figurepath}
			\else
				\@decmprtrue
				\def\@p@sfile{\ps@founddir\ps@dir#1}
				\def\@p@sbbfile{\ps@founddir\ps@dir#1.bb}
			\fi
		\else
			\closein1
			\@decmprtrue
			\def\@p@sfile{#1}
			\def\@p@sbbfile{#1.bb}
		\fi
	\else
		\def\@p@sfile{\ps@founddir\ps@dir#1}
		\def\@p@sbbfile{\ps@founddir\ps@dir#1}
	\fi
   \else
	\closein1
	\def\@p@sfile{#1}
	\def\@p@sbbfile{#1}
   \fi
}
\def\@p@@sfile#1{\@p@@sfigure{#1}}
\def\@p@@sbbllx#1{
		\@bbllxtrue
		\dimen100=#1
		\edef\@p@sbbllx{\number\dimen100}
}
\def\@p@@sbblly#1{
		\@bbllytrue
		\dimen100=#1
		\edef\@p@sbblly{\number\dimen100}
}
\def\@p@@sbburx#1{
		\@bburxtrue
		\dimen100=#1
		\edef\@p@sbburx{\number\dimen100}
}
\def\@p@@sbbury#1{
		\@bburytrue
		\dimen100=#1
		\edef\@p@sbbury{\number\dimen100}
}
\def\@p@@sheight#1{
		\@heighttrue
		\dimen100=#1
   		\edef\@p@sheight{\number\dimen100}
}
\def\@p@@swidth#1{
		\@widthtrue
		\dimen100=#1
		\edef\@p@swidth{\number\dimen100}
}
\def\@p@@srheight#1{
		\@rheighttrue
		\dimen100=#1
		\edef\@p@srheight{\number\dimen100}
}
\def\@p@@srwidth#1{
		\@rwidthtrue
		\dimen100=#1
		\edef\@p@srwidth{\number\dimen100}
}
\def\@p@@sangle#1{
		\@angletrue
		\edef\@p@sangle{#1} 
}
\def\@p@@ssilent#1{ 
		\@verbosefalse
}
\def\@p@@sprolog#1{\@prologfiletrue\def\@prologfileval{#1}}
\def\@p@@spostlog#1{\@postlogfiletrue\def\@postlogfileval{#1}}
\def\@cs@name#1{\csname #1\endcsname}
\def\@setparms#1=#2,{\@cs@name{@p@@s#1}{#2}}
\def\ps@init@parms{
		\@bbllxfalse \@bbllyfalse
		\@bburxfalse \@bburyfalse
		\@heightfalse \@widthfalse
		\@rheightfalse \@rwidthfalse
		\def\@p@sbbllx{}\def\@p@sbblly{}
		\def\@p@sbburx{}\def\@p@sbbury{}
		\def\@p@sheight{}\def\@p@swidth{}
		\def\@p@srheight{}\def\@p@srwidth{}
		\def\@p@sangle{0}
		\def\@p@sfile{} \def\@p@sbbfile{}
		\def\@p@scost{10}
		\def\@sc{}
		\@prologfilefalse
		\@postlogfilefalse
		\@clipfalse
		\if@noisy
			\@verbosetrue
		\else
			\@verbosefalse
		\fi
}
\def\parse@ps@parms#1{
	 	\@psdo\@psfiga:=#1\do
		   {\expandafter\@setparms\@psfiga,}}
\newif\ifno@bb
\def\bb@missing{
	\if@verbose{
		\ps@typeout{psfig: searching \@p@sbbfile \space  for bounding box}
	}\fi
	\no@bbtrue
	\epsf@getbb{\@p@sbbfile}
        \ifno@bb \else \bb@cull\epsf@llx\epsf@lly\epsf@urx\epsf@ury\fi
}	
\def\bb@cull#1#2#3#4{
	\dimen100=#1 bp\edef\@p@sbbllx{\number\dimen100}
	\dimen100=#2 bp\edef\@p@sbblly{\number\dimen100}
	\dimen100=#3 bp\edef\@p@sbburx{\number\dimen100}
	\dimen100=#4 bp\edef\@p@sbbury{\number\dimen100}
	\no@bbfalse
}
\newdimen\p@intvaluex
\newdimen\p@intvaluey
\def\rotate@#1#2{{\dimen0=#1 sp\dimen1=#2 sp
		  \global\p@intvaluex=\cosine\dimen0
		  \dimen3=\sine\dimen1
		  \global\advance\p@intvaluex by -\dimen3
		  \global\p@intvaluey=\sine\dimen0
		  \dimen3=\cosine\dimen1
		  \global\advance\p@intvaluey by \dimen3
		  }}
\def\compute@bb{
		\no@bbfalse
		\if@bbllx \else \no@bbtrue \fi
		\if@bblly \else \no@bbtrue \fi
		\if@bburx \else \no@bbtrue \fi
		\if@bbury \else \no@bbtrue \fi
		\ifno@bb \bb@missing \fi
		\ifno@bb \ps@typeout{FATAL ERROR: no bb supplied or found}
			\no-bb-error
		\fi
		\count203=\@p@sbburx
		\count204=\@p@sbbury
		\advance\count203 by -\@p@sbbllx
		\advance\count204 by -\@p@sbblly
		\edef\ps@bbw{\number\count203}
		\edef\ps@bbh{\number\count204}
		\if@angle 
			\Sine{\@p@sangle}\Cosine{\@p@sangle}
	        	{\dimen100=\maxdimen\xdef\r@p@sbbllx{\number\dimen100}
					    \xdef\r@p@sbblly{\number\dimen100}
			                    \xdef\r@p@sbburx{-\number\dimen100}
					    \xdef\r@p@sbbury{-\number\dimen100}}
                        \def\minmaxtest{
			   \ifnum\number\p@intvaluex<\r@p@sbbllx
			      \xdef\r@p@sbbllx{\number\p@intvaluex}\fi
			   \ifnum\number\p@intvaluex>\r@p@sbburx
			      \xdef\r@p@sbburx{\number\p@intvaluex}\fi
			   \ifnum\number\p@intvaluey<\r@p@sbblly
			      \xdef\r@p@sbblly{\number\p@intvaluey}\fi
			   \ifnum\number\p@intvaluey>\r@p@sbbury
			      \xdef\r@p@sbbury{\number\p@intvaluey}\fi
			   }
			\rotate@{\@p@sbbllx}{\@p@sbblly}
			\minmaxtest
			\rotate@{\@p@sbbllx}{\@p@sbbury}
			\minmaxtest
			\rotate@{\@p@sbburx}{\@p@sbblly}
			\minmaxtest
			\rotate@{\@p@sbburx}{\@p@sbbury}
			\minmaxtest
			\edef\@p@sbbllx{\r@p@sbbllx}\edef\@p@sbblly{\r@p@sbblly}
			\edef\@p@sbburx{\r@p@sbburx}\edef\@p@sbbury{\r@p@sbbury}
		\fi
		\count203=\@p@sbburx
		\count204=\@p@sbbury
		\advance\count203 by -\@p@sbbllx
		\advance\count204 by -\@p@sbblly
		\edef\@bbw{\number\count203}
		\edef\@bbh{\number\count204}
}
\def\in@hundreds#1#2#3{\count240=#2 \count241=#3
		     \count100=\count240	
		     \divide\count100 by \count241
		     \count101=\count100
		     \multiply\count101 by \count241
		     \advance\count240 by -\count101
		     \multiply\count240 by 10
		     \count101=\count240	
		     \divide\count101 by \count241
		     \count102=\count101
		     \multiply\count102 by \count241
		     \advance\count240 by -\count102
		     \multiply\count240 by 10
		     \count102=\count240	
		     \divide\count102 by \count241
		     \count200=#1\count205=0
		     \count201=\count200
			\multiply\count201 by \count100
		 	\advance\count205 by \count201
		     \count201=\count200
			\divide\count201 by 10
			\multiply\count201 by \count101
			\advance\count205 by \count201
		     \count201=\count200
			\divide\count201 by 100
			\multiply\count201 by \count102
			\advance\count205 by \count201
		     \edef\@result{\number\count205}
}
\def\compute@wfromh{
		\in@hundreds{\@p@sheight}{\@bbw}{\@bbh}
		\edef\@p@swidth{\@result}
}
\def\compute@hfromw{
	        \in@hundreds{\@p@swidth}{\@bbh}{\@bbw}
		\edef\@p@sheight{\@result}
}
\def\compute@handw{
		\if@height 
			\if@width
			\else
				\compute@wfromh
			\fi
		\else 
			\if@width
				\compute@hfromw
			\else
				\edef\@p@sheight{\@bbh}
				\edef\@p@swidth{\@bbw}
			\fi
		\fi
}
\def\compute@resv{
		\if@rheight \else \edef\@p@srheight{\@p@sheight} \fi
		\if@rwidth \else \edef\@p@srwidth{\@p@swidth} \fi
}
\def\compute@sizes{
	\compute@bb
	\if@scalefirst\if@angle
	\if@width
	   \in@hundreds{\@p@swidth}{\@bbw}{\ps@bbw}
	   \edef\@p@swidth{\@result}
	\fi
	\if@height
	   \in@hundreds{\@p@sheight}{\@bbh}{\ps@bbh}
	   \edef\@p@sheight{\@result}
	\fi
	\fi\fi
	\compute@handw
	\compute@resv}
\def\OzTeXSpecials{
	\special{empty.ps /@isp {true} def}
	\special{empty.ps \@p@swidth \space \@p@sheight \space
			\@p@sbbllx \space \@p@sbblly \space
			\@p@sbburx \space \@p@sbbury \space
			startTexFig \space }
	\if@clip{
		\if@verbose{
			\ps@typeout{(clip)}
		}\fi
		\special{empty.ps doclip \space }
	}\fi
	\if@angle{
		\if@verbose{
			\ps@typeout{(rotate)}
		}\fi
		\special {empty.ps \@p@sangle \space rotate \space} 
	}\fi
	\if@prologfile
	    \special{\@prologfileval \space } \fi
	\if@decmpr{
		\if@verbose{
			\ps@typeout{psfig: Compression not available
			in OzTeX version \space }
		}\fi
	}\else{
		\if@verbose{
			\ps@typeout{psfig: including \@p@sfile \space }
		}\fi
		\special{epsf=\ps@predir\@p@sfile \space }
	}\fi
	\if@postlogfile
	    \special{\@postlogfileval \space } \fi
	\special{empty.ps /@isp {false} def}
}
\def\DvipsSpecials{
	\special{ps::[begin] 	\@p@swidth \space \@p@sheight \space
			\@p@sbbllx \space \@p@sbblly \space
			\@p@sbburx \space \@p@sbbury \space
			startTexFig \space }
	\if@clip{
		\if@verbose{
			\ps@typeout{(clip)}
		}\fi
		\special{ps:: doclip \space }
	}\fi
	\if@angle
		\if@verbose{
			\ps@typeout{(clip)}
		}\fi
		\special {ps:: \@p@sangle \space rotate \space} 
	\fi
	\if@prologfile
	    \special{ps: plotfile \@prologfileval \space } \fi
	\if@decmpr{
		\if@verbose{
			\ps@typeout{psfig: including \@p@sfile.Z \space }
		}\fi
		\special{ps: plotfile "`zcat \@p@sfile.Z" \space }
	}\else{
		\if@verbose{
			\ps@typeout{psfig: including \@p@sfile \space }
		}\fi
		\special{ps: plotfile \@p@sfile \space }
	}\fi
	\if@postlogfile
	    \special{ps: plotfile \@postlogfileval \space } \fi
	\special{ps::[end] endTexFig \space }
}
\def\psfig#1{\vbox {
	\ps@init@parms
	\parse@ps@parms{#1}
	\compute@sizes
	\ifnum\@p@scost<\@psdraft{
		\PsfigSpecials 
		\vbox to \@p@srheight sp{
			\hbox to \@p@srwidth sp{
				\hss
			}
		\vss
		}
	}\else{
		\if@draftbox{		
			\hbox{\fbox{\vbox to \@p@srheight sp{
			\vss
			\hbox to \@p@srwidth sp{ \hss 
			 \hss }
			\vss
			}}}
		}\else{
			\vbox to \@p@srheight sp{
			\vss
			\hbox to \@p@srwidth sp{\hss}
			\vss
			}
		}\fi

	}\fi
}}
\psfigRestoreAt
\setDriver
\let\@=\LaTeXAtSign

\markboth{H\"oflich\& Wheeler }{SPECTRAL ANALYSIS OF SN1987A}
\setcounter{page}{1}

\begin{document}

\title{Spectral Analysis of SN1987A}

\author{P. H\"oflich and J. C. Wheeler}
\affil{Astronomy Department, University of Texas, Austin, TX 78712}

\begin{abstract}
 We examine the current status of the
spectral analysis of SN1987A during its early stages. 
Issues of the shock breakout and UV flash, the density 
and chemical structure, and masses of  different layers are discussed. 
A decade later, several aspects need a fresh look and interpretation. 
We summarize what questions have been answered and where some
results disagree.  Unresolved problems such as the excess of s-process
elements and influence of asphericity are addressed.
Finally, SN1987A is considered as a test case for using Type II as 
distance indicators via the Baade-Wesselink method.
\end{abstract}


\section{Introduction}
The emitted light of Type II supernovae allows study of
the explosion mechanism, testing of hydrodynamic models and probing of
the final stages of the stellar evolution.
The observed spectrum gives direct information on the
physical and chemical conditions of the supernova photosphere
at a given time. Deeper layers are observable at later times. 
Spectral analysis provides an effective tool to scan through the 
atmosphere, to reveal the density, chemical, and velocity structure, 
to probe explosion models, to reveal mixing processes and departures 
from sphericity, and to determine the distance.
At any one time, the photosphere spans only about 
$10^{-4}$ to $10^{-2} M_\odot$ and, consequently, 
good time coverage is critical to derive integrated quantities such 
as the total mass.
The following analysis is mainly based on our own work but, for comparison,
we will refer to results of other groups as appropriate (Table 1). 
Specific references are given in the text.
 
 \begin{table}
 \caption{Some Work on Atmosphere Models of SN 1987A}
 \begin{center}\scriptsize
 \begin{tabular}{crrrrrrrrrrr}
 
    First Authors~~~~~~~~~~~~~ &~~Phase ~ & ~~~~~~~~~~~~~~~~~~~~~~ Topic 
 ~~~~~~~~~~~~~~~~~~~~~~~~~~~~~~~~   \\
 \tableline
 \tableline
    Ensman/Hauschildt & 0-4 d  & shock breakout, early spectra + colors      \\
 \tableline
    H\"oflich           & 0-220d &  shock breakout, abundances, structure,      \\
                        &        & mixing, asphericities, CO, distance \\
 \tableline
    Lucy/Mazzali        & 2-100d &  structure + abundances                      \\
 \tableline
    Jeffery             & 2-30d   &  asphericity                                 \\
 \tableline
    Eastman            & 2-10d   &  structure/distance by Baade-Wesselink     \\
 \tableline
    Schwarz             & 150-500d &  gamma excitation                         \\
 \tableline
    Li/McCray           & 200+d &  gamma excitation, mixing                 \\
 \tableline
    Wagoner/Chilukuri   & 2-50 d&  distance by Baade-Wesselink              \\
 \tableline
 \end{tabular}
 \end{center}
 \end{table}
 
\section {Model Assumptions and Free Parameters}
 
 Most of the calculations summarized here were 
performed prior to 1989 and, consequently,
the description of our Nlte code for Extended ATmospheres (NEAT)
below corresponds to this era
(H\"oflich 88ab, 90, 91a; hereafter H stands for H\"oflich).
It has since been updated. 
The system of equations and their solution is briefly described
as follows. a) The radiation transport equation is solved  
in the comoving frame and includes relativistic velocity fields 
and line blanketing (Mihalas et al., 1975, 76ab).
Time dependent terms can be neglected because the radiative time 
scales are short compared to those of physical quantities at the photosphere.
b) The statistical equations are solved by a perturbation method (ALI).
  For the early light curve (LC), 
time dependent statistical equations are solved (H91a).
c) The energy equation. 
During later phases, we assume homologous expansion
and radiative equilibrium to determine the thermal structure.
 The bolometric and $\gamma$-ray luminosities are taken from
the observations.  The $\gamma$-ray energy deposition is treated 
following Colgate et al. (1980).  During the shock breakout,
the temperature structure is taken from the hydro because it is governed 
by adiabatic expansion and the corresponding recession in mass of
the photosphere (see also Hauschildt \& Ensman 1994, hereafter HE94).
Detailed atomic models are used for up to the three most abundant
ionization stages of several elements (H, He, C, N, O, Na, Mg, K, Ca).
 A large number of LTE-lines is included in a pure scattering approximation.
 
\section{The Shock Breakout}
 
 SN1987A is the first Type II supernova to have been 
observed shortly after the initial
event (e.g. McNaught 1987).  Measurements of the envelope mass
and explosion energy based on     the early LC
are consistent with the interpretation of
the late LC  (Arnett, 1988; Nomoto et al., 1988;
Woosley, 1988); however, flux limited diffusion calculations
gave discrepancies between observations and
models of up to 2 magnitudes at the earliest phases.
 
\begin{figure}
\hskip 5.8cm
\psfig{figure=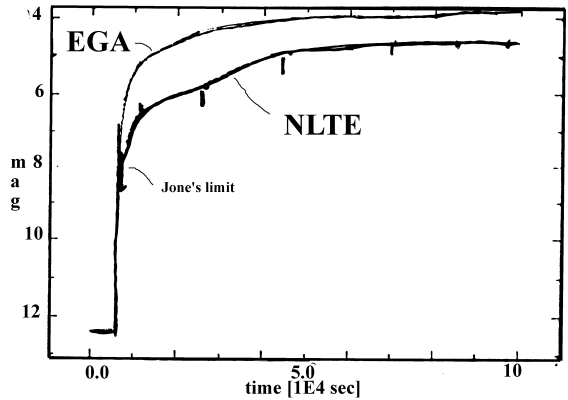,width=7.6cm,rwidth=5.5cm,clip=,angle=360}
\vskip -5.3cm
\caption{
\hsize=5.3cm
 Observed m$_V$ (with measure\-ment error bars)
 of SN1987A
in comparison with that predicted by a
  model with $1.25\times10^{51}$ erg
(Shigeyama \&  Nomoto, 1990)
 using NLTE and extended grey atmospheres (EGA).
The distance is taken to be 48 kpc. $E_{B-V}$ is set to $0.15^m$.}
\end{figure}

To address this problem, H91a coupled the hydrodynamical calculations of
Shigeyama \&  Nomoto (1990) with a detailed atmosphere. The progenitor has
a total mass of 16 $M_\odot $, 10.4 $M_\odot$ of H,  
a radius $R_{ph}=3\times10^{12}$cm.  
Explosions have been investigated with $E_{kin}$ of $1.0\times10^{51}$ and 
$1.25\times10^{51}$ erg which give a reasonable reproduction of the late 
light curve.  Both NLTE and scattering effects are 
important (Fig. 1, H\"oflich et al., 1986,  H90, H91a, Pizzochero, 1990).
The NLTE fluxes differ from those resulting from a flux 
limited diffusion approach or the solution for an extended gray 
atmosphere (EGA), and the slope of the LC changes.  In
particular, $L_{V}$ shows a strong local minimum shortly after
shock break out.  The general tendencies, and the drop in  $L_{V}$, 
can be understood as a consequence of the rapid drop in $\rho $ and T 
and the resulting size of NLTE effects and of the scattering optical 
depth (H91a).  

Table 2 gives the results of NLTE atmosphere models 
during the shock breakout phase;  the electron temperature T, the
radius of the photosphere R, the particle density N$_o$ (in cgs)
at the Thomson optical depth of  2/3, M$_{V}$, U-B, B-V and the
bolometric luminosity $L_{bol}$ based on the hydrodynamic calculation of
Shigeyama \& Nomoto (1990).  The dynamical model had logarithmic
zoning that resolved the photosphere at the epoch of the Jones' limit. 
The functions $C_{V}$ and $C_{bol}$ 
give the ratio of the visual and bolometric fluxes, respectively,
calculated by NLTE models to those of the extended grey atmosphere (EGA).
Observed colors are in good agreement
(t=$4.4\times10^4$ s B-V=-0.2 ... 0.0;     
at $9.83\times10^4$ s U-B=-0.836, B-V=0.085, Moreno \&  Walker 1987).
\begin{table}
\caption{Results of NLTE calculations of Shock Breakout} 
\begin{center}\scriptsize
\begin{tabular}{crrrrrrrrrrr}
time $\lbrack s \rbrack $
&~ T$ \lbrack K \rbrack$ &~ R~~ 
&~ N$_o$~~&~~ M$_V$&~U-B&~B-V&
~L$_{bol}$&~C$_V$&C$_{bol}$\\
\tableline
~6175.& 788000.&~3.36E12&~~6.3E14&
~~~-11.75&~~-1.23&~-0.30&~4.6E44&~0.18&0.39\\
~6400.&399450.&~4.20E12&~~7.8E13&
~~~-10.99&~~-1.28&~-0.32&~8.7E43&~0.13&0.32\\
~7000.&282000.&~6.28E12&~~3.2E13&
~~~-11.19&~~-1.33&~-0.33&~5.0E43&~0.12&0.28\\
~8000.&102000.&~9.56E12&~~1.1E13&
~~~-11.46&~~-1.23&~-0.28&~8.5E42&~0.19&0.24\\
10791.&~56200.&~1.90E13&~~5.2E11&
~~~-12.45&~~-1.20&~-0.29&~9.2E41&~0.21&0.22\\
26411.&~27000.&~5.44E13&~~1.8E11&
~~~-13.21&~~-1.02&~-0.24&~1.6E41&~0.23&0.30\\
44974.&~18700.&~8.92E13&~~9.8E10&
~~~-14.16&~~-0.98&~-0.22&~1.2E41&~0.30&0.36\\
70104.&~14900.&~1.28E14&~~7.5E10&
~~~-14.36&~~-0.95&~-0.14&~8.5E40&~0.33&0.38\\
99892.&~12100.&~1.70E14&~~5.0E10&
~~~-14.46&~~-0.82&~-0.07&~8.3E40&~0.37&0.43\\
\tableline
 
\end{tabular}
\end{center}
\end{table}

 The explosion energy $E_{kin}$ is not well determined 
by the absolute visual flux because of uncertainties in the 
distance and measurements, but can be estimated rather precisely by 
the rise time and the location of the first local minimum in V  
that depend on the shock travel time.
 Assuming that the time of the initial
core collapse is given by the neutrino detections, a value of 
$E_{kin}=1.0\times10^{51}$ erg can be ruled out.
Since the shock travel time through the progenitor
is essentially proportional to $E_{kin}^{0.5}$, the energy must 
be larger than $1.20\times10^{51}$ erg. 
A value $E_{kin}$ = $1.25\times10^{51}$ erg leads to a good agreement with
the observations.  Another, less stringent, upper limit  can be derived
from the upper limit set by Jones. SN1987A would have been seen
by Jones if the shock travel time were about 30 minutes shorter.
Therefore, $E_{kin}$ must be $\leq 1.6\times10^{51}$erg.
The hard UV fluxes are consistent with the 
observed high excitation lines of  NV and OVI
 in the circumstellar ring (Fransson \&  Lundqvist, 1989).
 
In the calculations just described no correction to the original 
temperature distribution given by flux-limited diffusion was
made for the NLTE radiation field.  
This limitation was addressed by Mair et al. (1992)
who used a gray radiation hydro code with adaptive mesh. For  Arnett's 
15 $M_\odot$ model, Mair et al. found that $E_{kin}$ must be close to
$1.3\times10^{51}$ erg and excluded $1.0\times10^{51}$ erg. 
Ensman \& Burrows (1992) and HE94 analyzed the explosion 
of Arnett's 17 $M_\odot $ model with an explosion energy
of $1.0\times10^{51}$ erg. These analyses are similar to the approach 
just described using a combination of a gray radiation hydro code and 
NLTE-atmospheres (without time dependence).  The explosion was
followed up to day 4. As also stated in HE94,
a problem seems to be that the photosphere recedes too rapidly and, 
consequently, the Doppler shifts of the absorption lines 
are too low after about day two. This suggests that a larger value of 
$E_{kin}$ is required, as concluded above.
 
\section{The Photospheric Phase}

A variety of structures for the later photospheric phase were tested 
(e.g. H87, H88ab), based on explosions of models with a main sequence  masses 
between 15 to 20 $M_\odot$ provided by Wei\ss (private 
communication). The best agreement was found for 18 $M_\odot$. 
A statistical velocity field v$_{stat}$  was introduced 
to improve the  agreement between observed and calculated
spectra. The size of v$_{stat}$ is small in comparison 
to v$_{exp}$ but larger than the sound speed,
$\approx 15 \% $ of v$_{exp}$ during the first week and 5 \% thereafter.
This quantity should not be interpreted as microturbulence, but
as a velocity field with a scale height smaller than the
free mean path of the line photons (H87,H88a, see \S6).
With increasing v$_{stat}$, the features become broader, 
the emission component is more red shifted and the rise from 
the minima to the maxima of the P-Cygni profile is slower (Fig. 2).

The luminosity was taken to be proportional to the observed 
bolometric LC.  The choice of $E_{kin}$ and the distance
then determines the time evolution of 
the photospheric parameters such as $R_{ph} $, $T_{eff}$, 
the radius $R_{HII} $ of the outer boundary of the ionized hydrogen 
region, and the matter velocity at the photosphere (see Fig. 3). 
The spectral evolution and the colors are well reproduced as
shown in Figures 4 and 5 (H87, H88ab).
Spectral analysis of the photospheric
phase gives $E_{kin}=1.3\times10^{51}$ erg, consistent with the 
shock break out and early LC analysis.
The distance is discussed in \S5.
 
\begin{figure}
\hskip 6.1cm
\psfig{figure=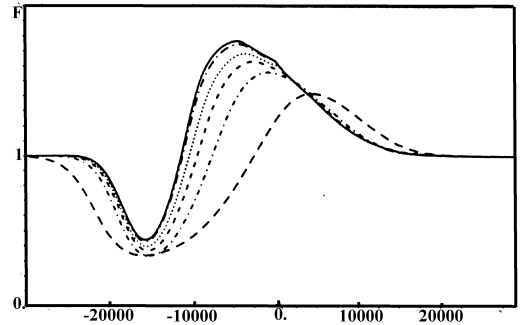,width=7.6cm,rwidth=5.5cm,clip=,angle=360}
\vskip -5.0cm
\caption{
\hsize=5.9cm
 $H_\alpha $ profiles with $v_{stat}/v(R_{ph})$ [in \%]  of 0 (solid), 1 (long dashed-
dotted), 5 (dotted), 10 (short dashed), 20 (short dashed dotted) and 50 (long dashed)
for a parameterized model with $T_{eff} = 10,000$ K,
$R_{ph}$ = 1.3$\times10^{15}$ cm, $v(R_{ph}$) = 10,000 km s$^{-1}$ 
and $\rho \propto r^{-7}$ (Duschinger et al. 1995).}
\end{figure}

\begin{figure}
\hskip 5.6cm
\psfig{figure=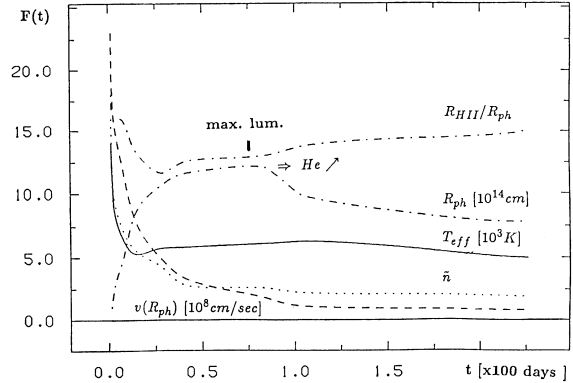,width=7.6cm,rwidth=5.5cm,clip=,angle=360}
\vskip -4.5cm
\caption{
\hsize=5.4cm
Distance $R_{ph}$,  $T_{eff}$, velocity v at $R_{ph}$, the power law
index $\tilde n$ of the corresponding power law density slope and the distance
$R_{HII} $ up to which hydrogen is mainly ionized are given as a function of time.}
\end{figure}

The difference  between the spectral evolution of SN1987A and 
most other SN II can be understood  as a consequence of the
compact structure of the progenitor star which implies higher densities and
steeper density gradients, and a low total luminosity.  For SN1987A,
the recombination phase of the photosphere is encountered
already after about 1 to 2 weeks.
 
Early on, the location of $R_{ph}$ is strongly coupled to the expanding 
matter due to the high temperature and density at the photosphere.
Consequently, the effective temperature $T_{eff}$ drops very
rapidly in the presence of a slowly varying total luminosity.  This causes
the rapid variations of the continuum flux
in the UV,  optical and  IR during the first week.
The smaller changes in the following months are due to the
slow increase of R$_{ph}$, due mainly to a geometrical
dilution effect and to the recombination of H outside
the radius R$_{HII}$. The decrease of R$_{ph}$ after
maximum light is caused by the decreasing luminosity and by the higher
helium abundances. Helium is mainly neutral and does
not contribute to the electron density at the photosphere.
Thus, the increase of the color temperature after maximum
light is mainly a result of  the increased helium abundance.
 The IR-excess can be understood as owing to free-free radiation and
extension effects ($\approx $ 10 to 20 \%).
 No additional dust component is needed
during the first few months at least in the nearer IR (H88ab).
 
 The observed frequency shifts of the absorption components
 of the Balmer lines are formed at different depths 
and hence probe the
v/$\rho$ structure (H91b). The good agreement between observed and calculated
line shifts confirms the density profiles $\rho(r)$.
  Lucy (1987) found similar density gradients from his analyses
which are mainly based on the UV spectra.  HE94 show fits of 
similar quality during the first 3 days with more self-consistent 
models (i.e. they calculate the luminosity from the radiation-hydro 
calculation). Their assumed $\rho(r)$
is consistent with H87, H88, H91ab.  At t=2d: $\rho (R_{ph}) \propto r^{-9}$ 
vs. $\propto  r^{-10}$ in our work.  Both in  H87 and HE94 
the $\rho $ gradient changes over the line forming region by $\approx $
20 \%.  Eastman \& Kirshner (1989) used $R_{ph}$,
$T_{eff}$ and $\rho(r)\propto r^{-\tilde n}$ as free parameters at each time. 
Although unrealistic, they used constant power law density profiles. 
They find gradients consistent with those given above 
($\tilde n $(t=2 d) = 9; $\tilde n $(t=10 d) = 5 to 7 vs. 10 and 7.2,
respectively in H87).

\begin{figure}
\hskip 5.4cm
 \psfig{figure=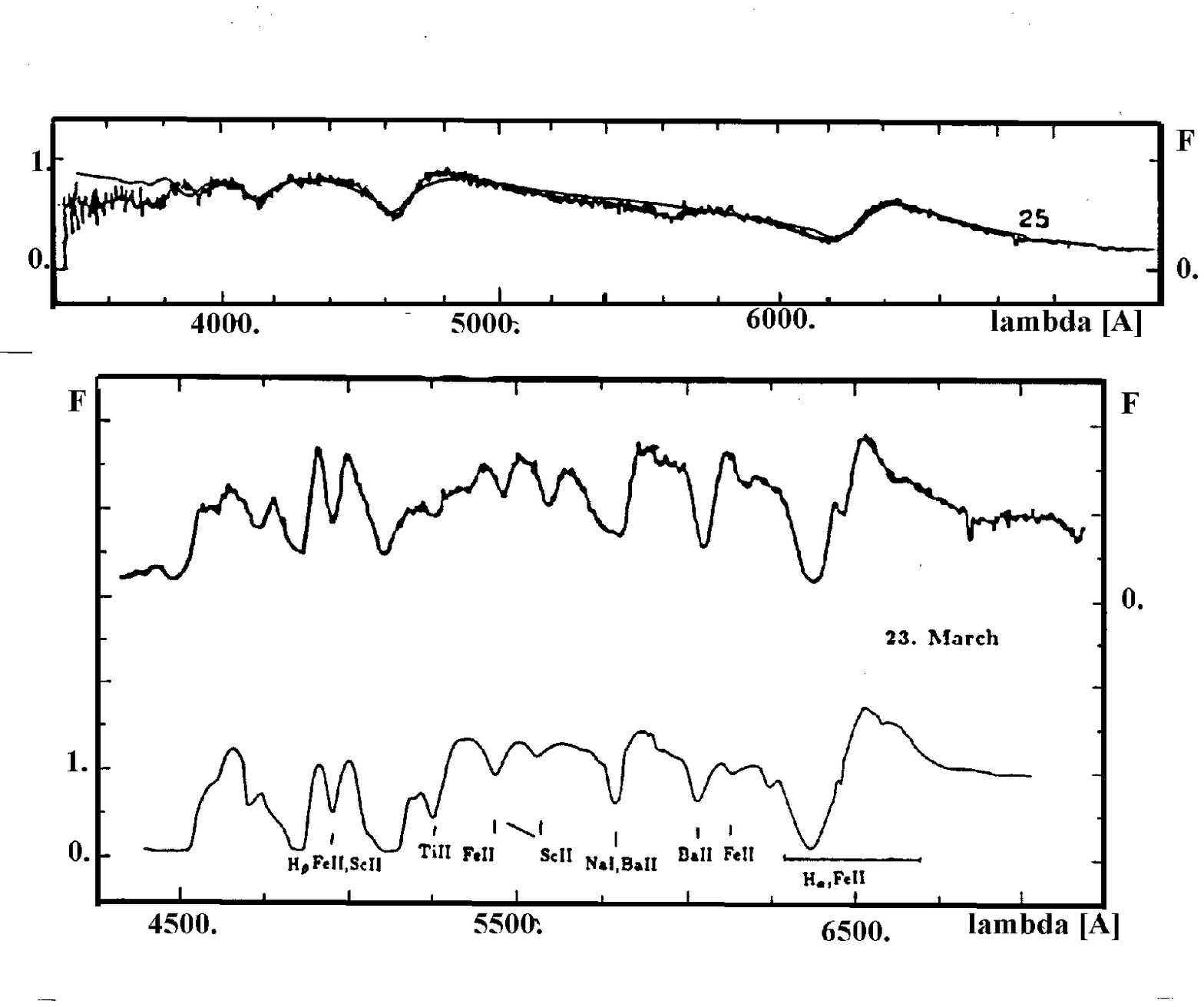,width=7.6cm,rwidth=5.5cm,clip=,angle=360}
\vskip -3.3cm
\caption{
\hsize=5.2cm
Spectra as observed by Menzies et al. (1987) 
 February 25.9 and March 23rd, 1987, in comparison with the
synthetic spectra.}
\end{figure}

After one to two weeks, the Thomson
scattering opacity declines due to recombination and heavy elements appear 
because weak lines are no longer ``washed out."
This spectral change does not imply a change in the composition.
Abundances less than solar are needed in order
to reduce line blanketing by the metal lines.
The abundances must be about one third of solar for nearly all elements (H88ab).
This value closely resembles that derived for the LMC by Dufour (1984).
The exception in SN 1987A is the s-process elements 
Sc, Ti, V, Cr, Sr, Ba and  Na. An overabundance relative to solar
of Sc, V, Cr, Sr  and Ti is needed to explain the slope of the
spectrum below 5300 \AA.  The strong line at about 
6100 \AA ~ can be attributed to Ba II (Williams  1988, H88ab) 
if Ba is increased by about 8-10 relative to LMC abundances.
Other BaII features are present in
the observed and calculated spectra at about 4900 \AA ~and 5000 \AA.
They are consistent with the same overabundance;
however blanketing by other lines prohibits  a detailed analysis.
Using Lucy's Monte Carlo code, Mazzali \&  Chugai (1995) 
reanalyzed the spectra and found an overabundance for Ba of about 
4 which is consistent within the uncertainties with other analyses.
 
%
%
%
%

The evolving spectra are 
well reproduced by the atmosphere models during the first 5 months;
however, later spectra (i.e. June, H88a) show insufficient
blue shifts indicating too small a photospheric velocity.
A power law density profile was introduced that yielded
higher central densities than the original dynamical model. 
This is a crude way to allow for the 
entropy redistribution that must accompany mixing.
At this epoch, excitation by $\gamma$-rays must also be taken into account 
because it strongly influences the temperature and ionization structure.
The size of the $\gamma$-ray flux is assumed to be
that given by the SMM satellite (Gehrels et al. 1989). 
$\gamma $-ray heating and inefficient cooling leads to high
temperatures in the outer H-envelope and substantial
emission by the cooling lines, e.g. Ca, despite their
low abundance (Schwarz 1991, Li \&  McCray 1996). 
 
\begin{figure}
\hskip 6.5cm
\psfig{figure=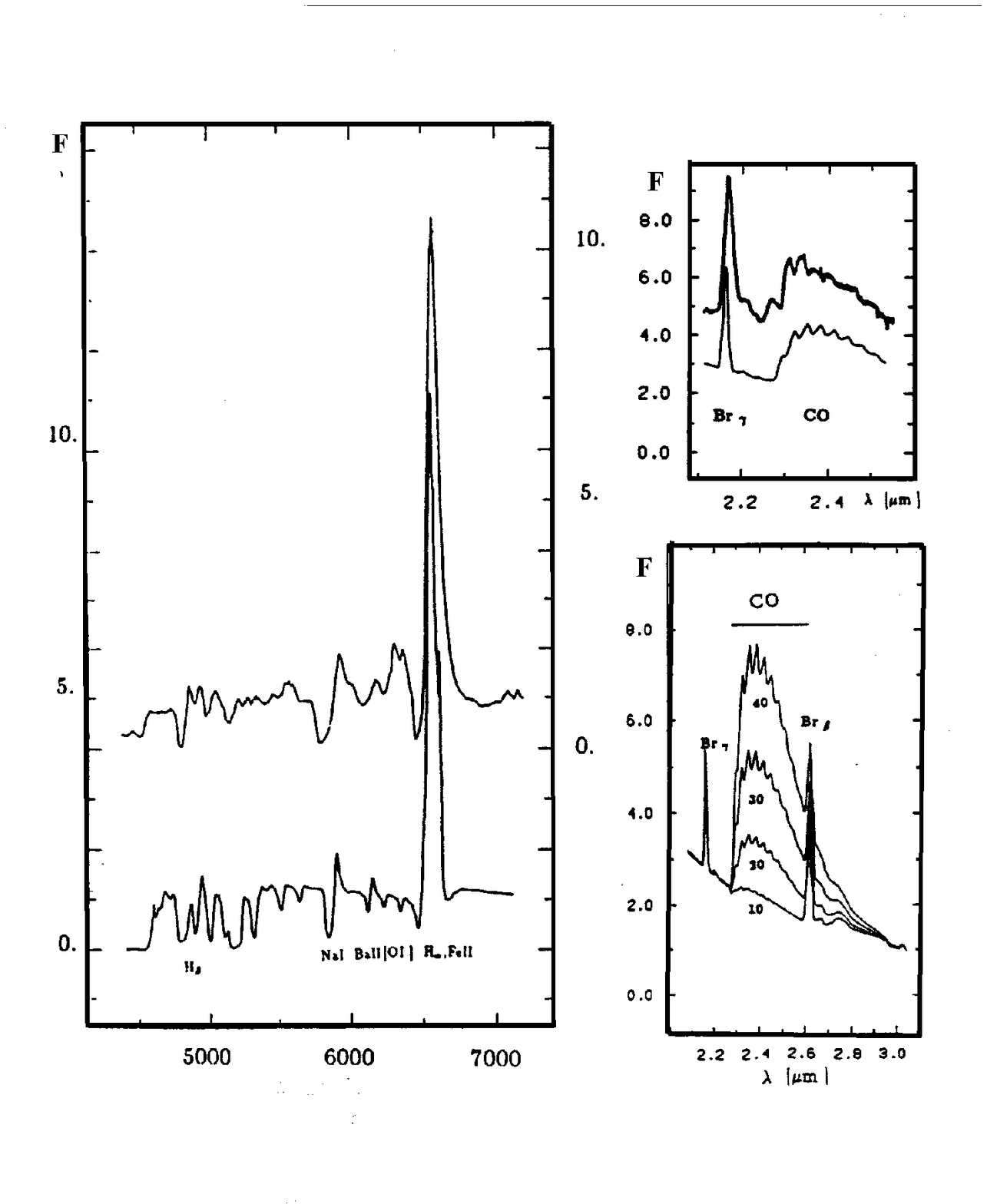,width=6.6cm,rwidth=6.5cm,clip=,angle=360}
\vskip -7.7cm
\hsize 7.3cm
\caption{
 Synthetic optical and IR spectra
 corresponding to October 2$^{nd}$, 1987, in comparison
with observations in the optical (Oct. 2$^{nd}$, Danziger, priv. comm.; left)
and in the IR (ESO observations between 1. and 6. Oct. 1987/ shifted by two units,
lower right). C and O was assumed to be enhanced to 20 * solar up to 
1800 km/sec. In addition, the theoretical CO spectra are given for various enhancement
factors (lower right).  Mixing out CO up to 3000 km/sec  would produce
featureless CO-bands (Sharp \& H\"oflich 1989). The feature at 2.38 $\mu $m cannot be attributed to $CO^+$.
}
\end{figure}
 The calculated and observed spectrum  on October 2 
(Fig. 5) are in good agreement, if we choose roughly
the same chemical abundances, but enrich He by a factor of 5.  
This implies that by this phase  He-rich layers are seen.
From the spectral analysis, the mass of the H-rich envelope 
is about 10 $M_\odot$ (H88ab).  A decrease of the H abundance is 
needed to explain the strong decrease of the continuum flux relative to 
the lines.  The increase in the He abundance leads to a strong 
decrease of R$_{ph}$, since  He is mainly neutral.
Furthermore, this explains in a consistent way
the increase of the color temperature given by  B-V.
At this time, hydrogen is still seen at the photosphere 
which shows an expansion velocity of the order of only 800 km s$^{-1}$. 
At the same time, CO bands in the IR
indicate that the matter is strongly enriched  with C and O
up to radii that correspond to
an expansion velocity of the order of 1800 km s$^{-1}$.
These results strongly indicate that about $3  M_{\odot}$ of H 
have been mixed into very deep layers (H88ab; Li et al. 1993) 
of the ejecta and
a moderate mass of C and O-rich matter, $\sim10^{-2}$ to $10^{-1} M_{\odot}$
have been mixed far out into the H-rich matter.  This  mixing
can be understood by Rayleigh-Taylor instabilities during the explosion.
  Mixing was also indicated by the
time dependence of the observed hard $\gamma $-rays, the
detection of broad IR lines of Co and Fe, and the
    LCs ( e.g. Woosley, 1988).

\section{Distance to SN1987A}

Distances to many SN~II have been derived by using the Baade-Wesselink method 
(e.g. SN 1969L: $12 \pm 4$ Mpc  and SN 1970G: $7 \pm 2$ Mpc  
Kirshner \&  Kwan, 1974; SN1979C: $23 \pm 3$ Mpc , Branch et al., 1981).
In this method the ratios of the observed and 
predicted fluxes and spectral slopes at different wavelengths 
together with the velocities at different times are used to derive 
quantities such as the photospheric radius.
In the past, these determinations
were based on the assumption that the radiation of the 
supernova can be represented by a black body; 
however, the dilution of the radiation field and line blanketing are critical 
(H\"oflich et al. 1986, Hershkowitz et al. 1986, H91c, Schmidt et al. 1994, 
Baron et al. 1995,  Eastman et al. 1996). For high temperatures, 
the dilution factors are rather insensitive to the model 
parameters (10 to 25 \%), but the absolute size given in the literature ranges
from $\approx 0.3 $ (Eastman et al.  1996) to 0.5 (Baron et al. 1995) 
with H91c being somewhere in between.  There is general agreement that
the dilution factor rises steeply with declining $T_{eff}$ during the
recombination phase of hydrogen (H91d, Schmidt et al. 1994). Typically, 
this phase is entered a few weeks after the explosion and is accompanied 
by the appearance of metal lines in the optical spectra.
The critical $T_{eff} $ for the onset of this rise
depends sensitively on several parameters such as 
the structure and the He/H ratio.
Unfortunately, uncertainties in the time since the 
explosion, $\Delta t$, enter the distance determination
as $\Delta t/t$.  This, together with the large sensitivity of the 
dilution factor as a function of T in the relevant range
$\sim 5000$ K, poses a significant limitation 
on the accuracy of the Baade-Wesselink method; however high accuracy can
be achieved if a detailed analysis of a broad range of data 
(including the time of the explosion) can be done. 
SN1987A can be regarded as the benchmark to demonstrate the ``ideal" case.
Based on the detailed spectral analysis for the first 200 days, 
(H88ab) derived a distance of $48 \pm 4$ kpc,
well within the error bars of other groups 
(e.g. Chilukuri \& Wagoner 1988: $43 \pm 4$  kpc;
Schmutz et al. 1990: 46 kpc; Eastman \& Kirshner 1989:$ 49 \pm 5$ kpc).

\section{Deviations from Sphericity}
 
For SN1987A, polarization $P$ of up to 0.5 \% was observed early on. 
The level of $P$ decreased rapidly over the next few weeks 
(e.g. Mendez et al. 1988).  It was concluded that
the envelope of SN1987A showed asphericities
of about 10 to 20 \% and that Thomson scattering was the mechanism to produce
$P$ (Jeffery,  1991, Mendez et al. 1988, H\"oflich et al. 1989,
H91b).  Whether $P$ is caused by rapid rotation 
of the progenitor (Steinmetz \& H\"oflich 1992), by an 
asphericity of the SN~II-explosion mechanism (Yamada \& Sato 1991), by an
asymmetric luminosity input (H95), or dust in the surroundings 
(Wang \& Wheeler 1996) is still a question under debate. 
There is growing evidence that, whatever the combination of
mechanisms, polarization is a rather common phenomenon among 
SNe~II (Wang et al. 1995).  The implications for our understanding 
of the explosion mechanism and progenitor evolution are
obvious.   There is also a direct impact on the use of SNe~II as 
distance indicators because aspherical configurations also result 
in aspherical luminosities. An appreciable systematic bias will enter 
in statistical analyses when many SNe~II are used to determine $H_0$ because
the directional redistribution of photons is different from the 
probability distribution of the inclination angle seen by an observer (H91c). 
For a more detailed discussion of $P$
see Wang, Wheeler \& H\"oflich (this volume).
 \begin{figure}[t]
\hskip 5.7cm
  \psfig{figure=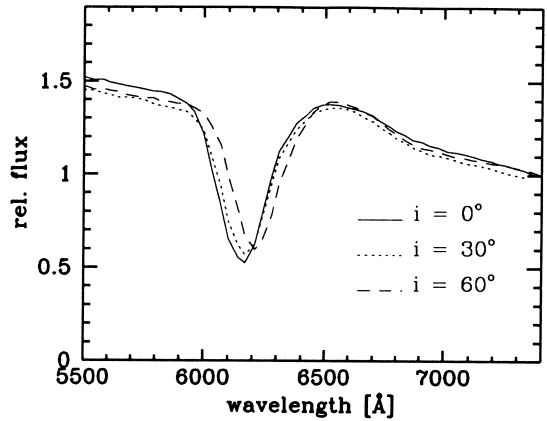,width=7.6cm,rwidth=5.5cm,clip=,angle=270}
\vskip -6.2cm
\caption{
\hsize=5.5cm
$H_\alpha $ profile for different inclinations $i$ for an oblate ellipsoid
calculated by a 3-D NLTE code. An axis ratio of 0.85 is assumed. Parameters 
representing the atmosphere of SN1987A at about day 3 after the explosion ($T_{eff}$= 9000 K,
$R_{ph,\tau_{sc}=1}=8.\times10^{14}$ cm, $\rho \propto r^{-9}$). For details, see
H\"oflich et al. (1995) \& Wang et al. (1997).
}
\end{figure}
 
The finite polarization of SN~II suggests a possible interpretation of 
the high statistical velocity deduced for SN 1987A as discussed in \S4.
An aspherical configuration has some qualitatively similar
effects on the spectra as a statistical velocity because both increase 
the spread in velocities observed at a given epoch, as can 
be seen by comparing Figures 2 and 6.  Both cause the rise from 
the minima to the maxima to be less steep and the emission is 
shifted compared to a spherical configuration without $v_{stat}$
(not shown in Fig. 6).
In spite of the similarity, there does not exist a direct 
correspondence between the parameters for a spherical configuration 
$v_{stat}$ and an aspherical configuration. More studies based on
3-D NLTE atmospheres are required (see Wang et al. 1997 for SNe~Ia).
 
\section {Conclusions and Open Questions}
 
 The peculiarities of the spectral and color evolution of 
SN1987A can be understood by the fact that the progenitor 
was a blue supergiant with its compact structure and low luminosity.
Both from the early LC and the spectra
we can conclude that  $E_{kin}$ is about $1.3\times10^{51}~\pm 0.3$ erg 
based on stellar models with $R_{ph}=3.10^{12}~cm$.  
The hydrogen-rich envelope is about $9-11 M_\odot$. 
The uncertainty of the total  mass of H is 
relatively small because of the weak dependence of the density
at the photosphere on the atmosphere model parameters.
Strong mixing of 3 $M_\odot $ of H with the inner layers
(down to 800 km s$^{-1}$) and mixing of C/O rich layers
out to about 1800 km s$^{-1}$ was found. This can be 
understood by Rayleigh-Taylor instabilities during the explosion,
but a full radiative transfer of a self-consistently mixed model
has not been done.
The metal abundances are consistent with the chemical composition of the LMC,
but the s-process elements are highly enriched. 
The abundance ratios of the s-process elements are in agreement 
with the s-process (Hashimoto et al. 1990), but the 
origin of the enhancement remains unexplained. 
Polarization suggests asphericity in the outer layers of SN1987A of about 
10 \%. This may explain the high statistical velocity field needed 
to fine tune spectral fits during early epochs (H87); however,
more detailed analysis is needed to test this suggestion.
 
This research is supported by NSF Grant AST9528110.


\begin{references}
\reference Arnett  W.D. 1988, ApJ 331, 377
\reference Baron E. et al. 1995, ApJ 441, 170
\reference Branch D. et al.  1981, ApJ 244,  780
\reference  Chilukuri  M., Wagoner  R.V., 1988, IAU Symposium 108, ed. K. Nomoto, Springer, p. 295
\reference Colgate S.A., Petschek A.G.,  Kreise J.T. 1980,   ApJ 237, L81
\reference Dufour  R.J. 1984, IAU Colloquium 108, ed. K. Nomoto, Springer, p. 353
\reference Duschinger  M., Puls J., Branch D., H\"oflich P., Gabler A. 1995 A\&A 297, 802
\reference Eastman R. G., Kirshner R.P. 1989, ApJ 347, 771
\reference Eastman R. G., Schmidt B.P., Kirshner R.P. 1996, ApJ 466, 911
\reference Ensman L., Burrows A. 1992, ApJ 393, 742
\reference Fransson  C., Lundqvist  P.  1988, ApJ 341, L59
\reference Gehrels,  N.,  MacCallum  C.J., Leventhal  M. 1989, ApJ 317, L73
\reference Hashimoto  M., Nomoto  K., Shigeyama  T. 1989, A\&A 210, L5
\reference    Hauschildt, P., Ensman L. 1994, ApJ 424, 905
\reference Hershkowitz  R., Lindner  E., Wagoner  B.  1986, A\&A 301, 220
\reference H\"oflich P., Wehrse R., Shaviv G. 1986, A\&A  163, 105
\reference H\"oflich P. 1987, in {\sl SN 1987A}, ed. by I.J. Danziger, ESO, p. 449 
\reference H\"oflich P. 1988a, IAU Colloquium 108., ed. K. Nomoto, Springer, p. 288
\reference  H\"oflich P. 1988b,  PASA 7, 434            
\reference H\"oflich P. 1990, Habil. Thesis, U. of Munich, MPA-563
\reference H\"oflich  P. 1991a, in {\sl SN 1987A \& other SNe},  ed. J. Danziger,
ESO, p. 387
\reference H\"oflich P. 1991b, A\&A 246, 481
\reference H\"oflich P. 1991c, in ``Supernovae", ed. S.E. Woosley, Springer,  p.415
\reference H\"oflich P. et al. 1995, ApJ   459, 307
\reference Jeffery D. 1991, ApJ 352, 267
\reference  Kirshner R.,  Kwan J., 1974,   ApJ  193,  27
\reference Li, H., McCray, R. A., and Sunayev, R. A. 1993, ApJ 419, 824
\reference Li, H., McCray, R. 1996 ApJ 456, 370
\reference Lucy, 1987, in  SN1987A \& other SNe, ed. J. Danziger, ESO, p. 302
\reference Mair, G., Hillebrandt, W., H\"oflich, P., Dorfi, E. A\&A 266, 266
\reference Mazzali M., Chugai, N.  1995 A\&A 303, 118
\reference McNaught  R. H., 1987  IAU Circular, 4389
\reference Mendez M., Clocchiatti A., Benvenuto  G., Feinstein  C. 1988, ApJ 334, 295
\reference Mihalas D., Kunasz R.B., Hummer D.G.  1975, ApJ 202, 465
\reference Mihalas D., Kunasz R.B., Hummer D.G.  1976a, ApJ 206, 515
\reference Mihalas D., Kunasz R.B., Hummer D.G.  1976b, ApJ 210, 419
\reference Moreno B., Walker S., 1987 IAU Circular 4316
\reference  Nomoto  K., Shigeyama  T., Kumagai  S., Hashimoto  S., 1988, PASA 7, 490
\reference Pizzochero P. 1991,  in SN1987A \& other SNe,  ed. J. Danziger,
ESO, p. 203
\reference Schmidt B.P. et al. 1994, ApJ 432, 42
\reference Schmutz W. et al 1990, ApJ 355, 255
\reference Schwarz D. 1991, in ``Supernovae", ed. S.E. Woosley,  Springer, p. 437
\reference Sharp C., H\"oflich P. 1989, Astr.Sp.Sc. 171, 213
\reference Shigeyama T., Nomoto K. 1990, ApJ 360, 242
\reference Steinmetz M., H\"oflich P. 1991, A\&A 257,  641
\reference Wang  L., Wheeler  J.C., Li, Z.W., Clocchiatti, A. 1996, ApJ 467, 435
\reference Wang  L.,  Wheeler  J.C. 1996, ApJ, 462, L27
\reference Wang  L., Wheeler  J.C.,  H\"oflich, P. 1997, ApJ 476, L27
\reference Williams  R.E. 1988, 4th George Mason Conf., ed.M. Kafatos, p. 355
\reference Woosley  S.E.  1988, IAU Colloquium 108, ed. K. Nomoto, Springer, p. 361
\reference Yamada, S.; Sato, K., 1990, ApJ 358, L9
\end{references}
\end{document}